\documentclass{ws-ijmpb}

\usepackage{amssymb,amsmath,graphicx,setspace,multirow,color,rotating,subfigure,url}
\usepackage{extarrows}
\usepackage{mathrsfs}
\usepackage{textcomp}
\usepackage{float}

\begin{document}

\markboth{H.-C. Xu, Z.-Q. Jiang and W.-X. Zhou}
{Immediate price impact of a stock and its warrant}

%
\catchline{}{}{}{}{}
%

\title{Immediate price impact of a stock and its warrant: \\
    Power-law or logarithmic model?}

\author{Hai-Chuan Xu}

\address{Research Center for Econophysics, Department of Finance and Postdoctoral Research Station,\\
 East China University of Science and Technology, \\
 130 Meilong Road, Shanghai 200237, P. R. China\\
 hcxu@ecust.edu.cn}

\author{Zhi-Qiang Jiang}

\address{Research Center for Econophysics and Department of Finance,\\
 East China University of Science and Technology,\\
 130 Meilong Road, Shanghai 200237, P. R. China\\
 zqjiang@ecust.edu.cn}

\author{Wei-Xing Zhou\footnote{Corresponding author.}}

\address{Research Center for Econophysics, Department of Finance and Department of Mathematics,\\
 East China University of Science and Technology,\\
 130 Meilong Road, Shanghai 200237, P. R. China\\
 wxzhou@ecust.edu.cn}

\maketitle

\begin{history}
\received{8 September 2016}
\revised{Day Month Year}
\accepted{(Day Month Year)}
\end{history}

\begin{abstract}
Based on the order flow data of a stock and its warrant, the immediate price impacts of market orders are estimated by two competitive models, the power-law model (PL model) and the logarithmic model (LG model). We find that the PL model is overwhelmingly superior to the LG model, regarding the robustness of the estimated parameters and the accuracy of out-of-sample forecasting. We also find that the price impacts of ask and bid orders are consistent with each other for filled trades, since significant positive correlations are observed between the model parameters of both types of orders. Our findings may provide valuable insights for optimal trade execution.
\end{abstract}

\keywords{Econophysics; immediate price impact; limit order book.}

PACS numbers: 89.65.Gh

\section{Introduction}
\label{S1:Intro}

The price impact of trade, a key indicator reflecting market liquidity, is of great importance for understanding price dynamics.\cite{Bouchaud-Farmer-Lillo-2009} Price impacts are found to be primarily dependent on signed trades' sizes.\cite{Chan-Fong-2000-JFE} In the original model of Kyle\cite{Kyle-1985-Em}, price impacts are assumed to be a linear function of order size. However, empirical studies show that the price impacts are a non-linear concave function of trade sizes. For example, power-law models with an exponent $\gamma$ ranging from 0.2 to 0.8 are uncovered from different datasets in terms of markets, periods and order properties \cite{Plerou-Gopikrishnan-Gabaix-Stanley-2002-PRE,Lillo-Farmer-Mantegna-2003-Nature,Farmer-Lillo-2004-QF,Moro-Vicente-Moyano-Gerig-Farmer-Vaglica-Lillo-Mantegna-2009-PRE,Almgren-Thum-Hauptmann-Li-2005-RISK,Lim-Coggins-2005-QF,Engle-Ferstenberg-Russell-2006-JPM,Gabaix-Gopikrishnan-Plerou-Stanley-2008-JEDC,Zhou-2012-QF}.
Several theoretical explanations are proposed for the size dependence of price impact. Such power-law behaviors will arise when the trading is performed in an optimal way. \cite{Gabaix-Gopikrishnan-Plerou-Stanley-2003-Nature,Gabaix-Gopikrishnan-Plerou-Stanley-2006-QJE} This anomalous impact has also been explained through agent-based simulations, mean-field theory or reaction-diffusion models.\cite{Daniels-Farmer-Gillemot-Iori-Smith-2003-PRL,Farmer-Patelli-Zovko-2005-PNAS,Toth-Lemperiere-Deremble-deLataillade-Kockelkoren-Bouchaud-2011-PRX,Mastromatteo-Toth-Bouchaud-2014-PRE,Mastromatteo-Toth-Bouchaud-2014-PRL,Mastromatteo-2014-JSM,Donier-Bonart-Mastromatteo-Bouchaud-2014-QF} Take Ref.~\refcite{Toth-Lemperiere-Deremble-deLataillade-Kockelkoren-Bouchaud-2011-PRX} for example, under the assumption of a linear latent order book, high participation rates of large orders can result in the power-law function for price impact.

Different with the power-law model, a logarithmic price impact relation for several stocks listed on the Paris Bourse has been proposed.\cite{Potters-Bouchaud-2003-PA} Similarly, Zarinelli et al. present that a logarithmic functional form fits large orders better than a power law does.\cite{Zarinelli-Treccani-Farmer-Lillo-2015-MML} However, they do not examine the time evolution of estimated parameters and do not carry out out-of-sample tests.

In this letter we aim to find which model is better in capturing the price impact of marketable orders by considering the stability of model parameters, the type of assets and the accuracy of out-of-sample prediction.

\section{Data description}

Our data contain the detail transaction records when marketable orders match the orders at the opposite of limit order book, which allow us to get the information of both matched orders and their transactions. Stock Baosteel (600019) and its warrant (580000), traded on Shanghai Stock Exchange, are included in our analysis. Our datasets cover the periods from 22 August, 2005 to 23 August, 2006. In the Shanghai Stock Exchange, the tick size of stock is 0.01 \textyen~(CNY) and that of warrant is 0.001 \textyen~(CNY).

\begin{table}[!htb]
  \centering
  \tbl{Summary statistics of trades}
  {\begin{tabular}{crrrr}
  \hline\hline
  Group & Number & PI(bps) & Volume & \textyen~Volume \\
  \hline
  \multicolumn{5}{l}{\textit{Stock}}\\
  FB & 429185 & 0.1727 & 10667.7 & 45345.7 \\
  FS & 394438 & -0.2109 & 13080.0 & 55590.5 \\
  PB & 10745 & 23.908 & 61932.7 & 263739.6 \\
  PS & 10791 & -23.947 & 79265.2 & 335438.7 \\
  \hline
  \multicolumn{5}{l}{\textit{Warrant}}\\
  FB & 3163177 & 1.3689 & 18837.7 & 24387.4 \\
  FS & 3056320 & -1.7178 & 20476.1 & 26227.6 \\
  PB & 161521 & 14.923 & 56337.5 & 68415.3 \\
  PS & 158699 & -15.260 & 52213.8 & 64136.5 \\
  \hline\hline
\end{tabular}}
\label{Tb:descriptive:statistics}
\end{table}

Following Ref.~\refcite{Biais-Hillion-Spatt-1995-JF}, the trades are classified into four types according to their directions and aggressiveness: trades resulting from filled buy orders (FB), trades resulting from filled sell orders (FS), trades resulting from partially filled buy orders (PB), and trades resulting from partially filled sell orders (PS). Table~\ref{Tb:descriptive:statistics} shows the numbers, average price impacts (measured in basis points), average volumes, and average \textyen~volumes of the four types of trades. For each trade, its price impact is defined as the absolute logarithmic difference between the prices right before and after the trade. We find that, for both warrant and stock, there are more filled trades than partially filled trades, while the average volumes and price impacts of filled trades are much smaller than those of partially filled trades.\cite{Zhou-2012-QF,Zhou-2012-NJP} It is reasonable that large orders are comparatively few but can cause larger impacts.

\begin{figure}[!tb]
  \centering
  \includegraphics[width=0.45\linewidth]{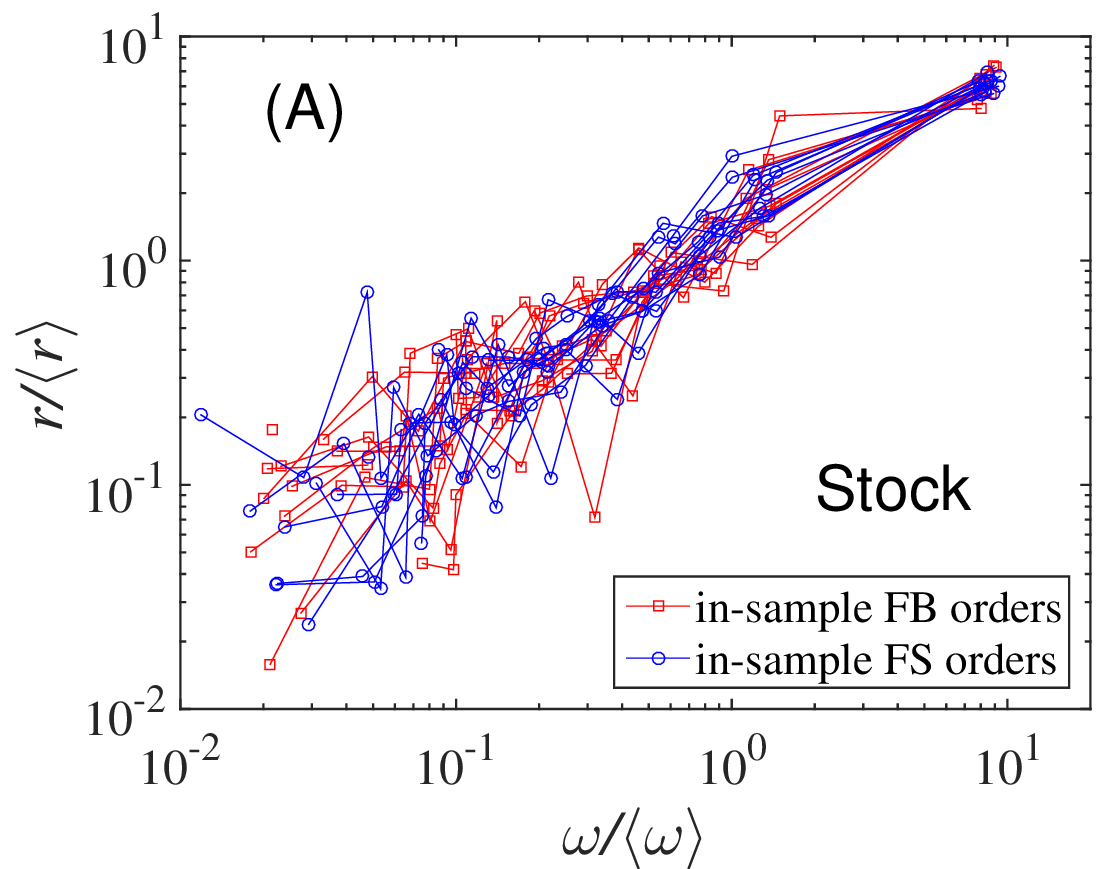}
  \includegraphics[width=0.45\linewidth]{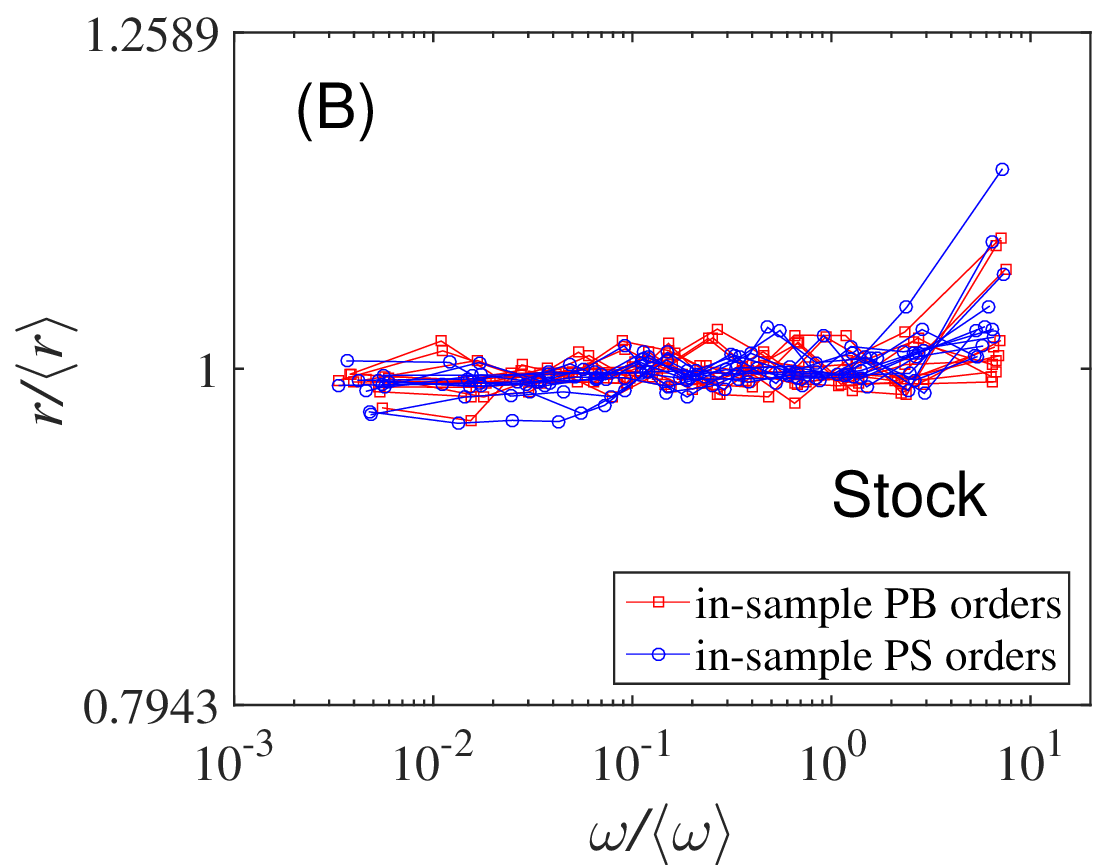}\\
  \vskip  +0.01\textwidth
  \includegraphics[width=0.45\linewidth]{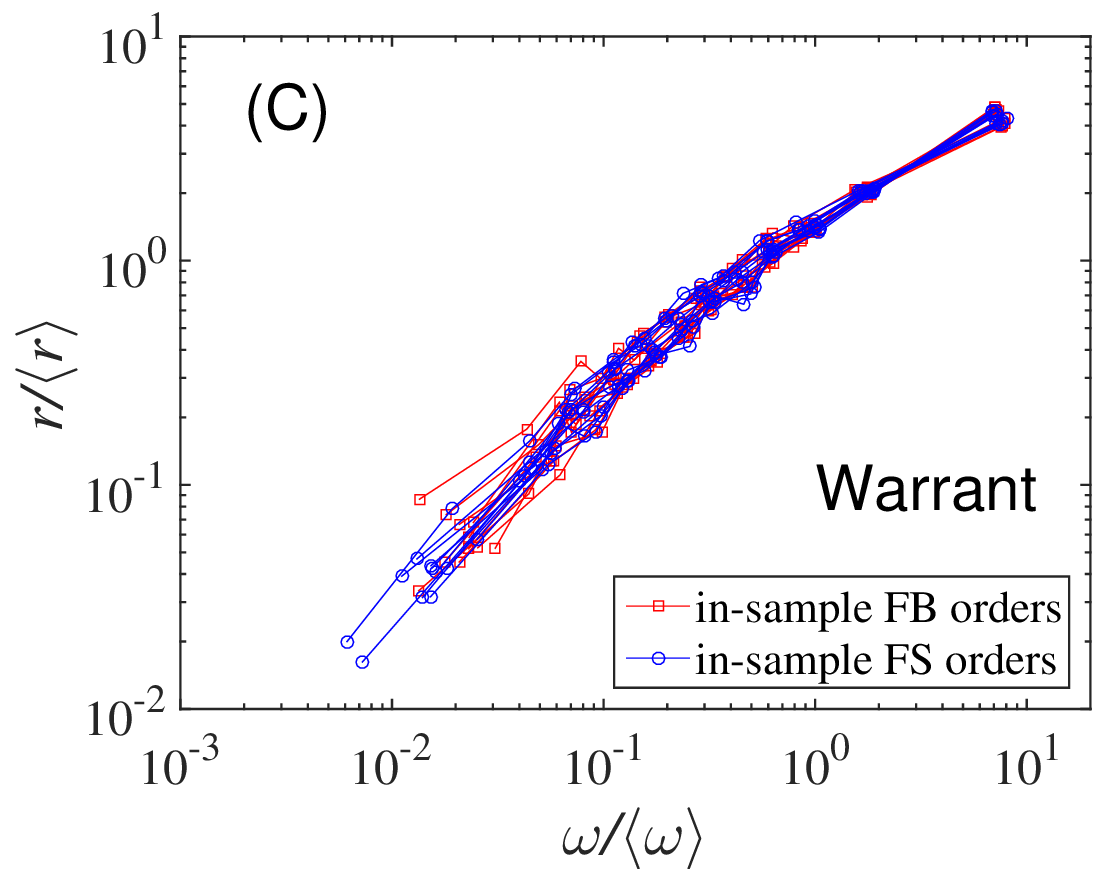}
  \includegraphics[width=0.45\linewidth]{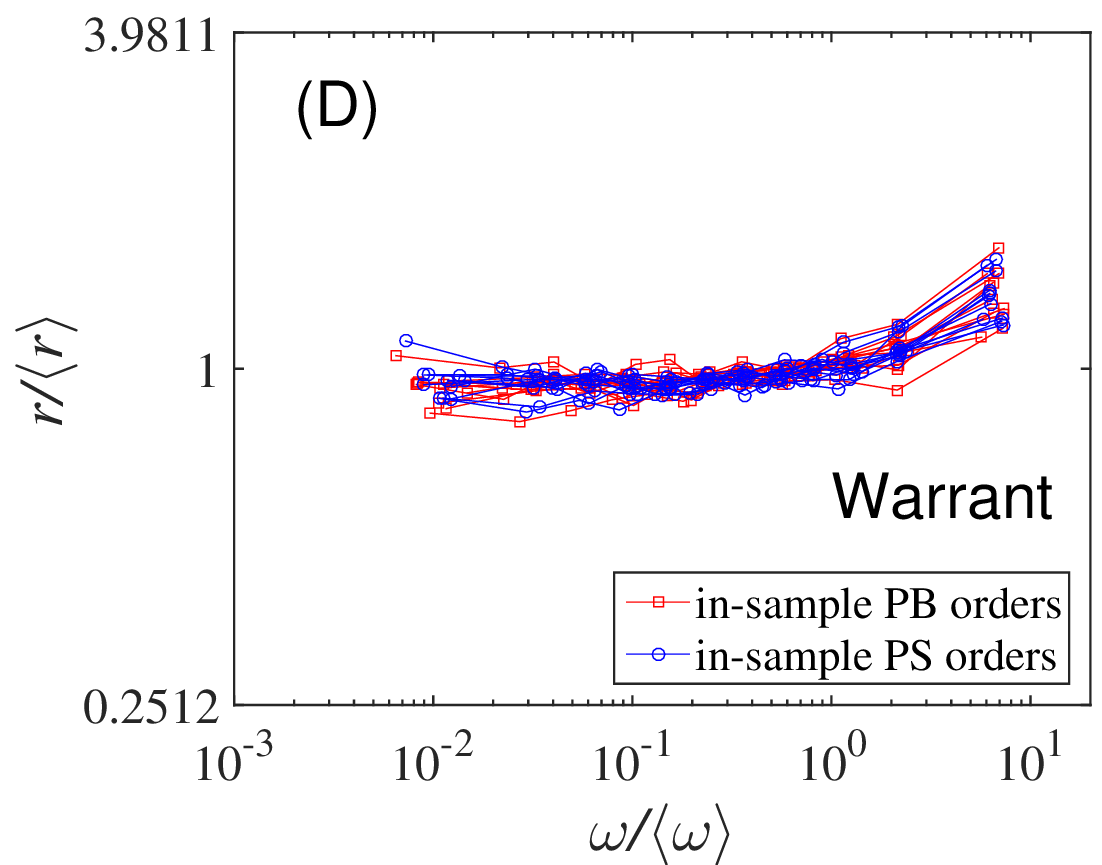}
  \caption{Immediate price impact for Stock 600019 and Warrant 580000 between August 2005 to August 2006. Each line corresponds to the average price impact of all relevant trades in time interval of one month. (A) Stock: filled trades. (B) Stock: partially filled trades. (C) Warrant: filled trades. (D) Warrant: partially filled trades.}\label{Fig:IPI:Insample}
\end{figure}

In order to study the stability of price impact, the whole data sample are divided into 12 segments and each segment contains the data of one month. In each month, the trades are equally split into $M=12$ groups according to their normalized trade size $\omega/\langle \omega \rangle$, which allow us to calculate the means of normalized price impact $r/\langle r \rangle$ and the means of normalized trade size $\omega/\langle \omega \rangle$ for each group. The resulting price impacts of each type of trades for the stock and the warrant are illustrated in Fig.~\ref{Fig:IPI:Insample}.

Several intriguing features are observed. First, the price impact patterns of filled trades (Fig.~\ref{Fig:IPI:Insample}(A) and Fig.~\ref{Fig:IPI:Insample}(C)) are significantly different from those of partially filled trades (Fig.~\ref{Fig:IPI:Insample}(B) and Fig.~\ref{Fig:IPI:Insample}(D)). In log-log scale coordinates, the price impact curves of the FB and FS trades can be roughly considered as linear or slightly concave with positive slopes, thus present power-law behaviors or logarithmic behaviors. However, for the PB and PS trades, $r/\langle r \rangle$ remains constant when $\omega/\langle \omega \rangle$ is not too large. Such difference exhibits in both stock (with large tick size) and warrant (with small tick size) and has been explained by Ref.~\refcite{Zhou-2012-QF}. Second, the lines of different months overlap together, indicating that the normalized price impacts are relatively stable. In addition, there is no significant asymmetry between the buyer-initiated and seller-initiated trades at the transaction level.

\section{Immediate price impact model}

Following above analysis, we will examine whether the power-law model or the logarithmic model fits the price impact better. Here we only consider the price impact of filled trades, due to the fact that the price impact of partially filled trades is roughly constant.

The power-law model (PL model) was proposed in Ref.~\refcite{Zhou-2012-QF}:
\begin{equation}\label{Eq:powerlaw:model}
  r/\langle r \rangle = a(\omega/\langle \omega \rangle)^{\gamma} + \varepsilon
\end{equation}
where $a$ and $\gamma$ are parameters to be estimated and $\varepsilon$ is an independent error term.

The logarithmic model (LG model) was proposed in Ref.~\refcite{Zarinelli-Treccani-Farmer-Lillo-2015-MML}:
\begin{equation}\label{Eq:logarithmic:model}
  r/\langle r \rangle = c\log_{10}(1+d(\omega/\langle \omega \rangle)) + \varepsilon
\end{equation}
where $c$ and $d$ are parameters to be estimated.

We estimate the model parameters with the trades in the first 11 months and leave the trades in the last month for out-of-sample tests. The nonlinear least square estimation is applied to both models. The estimation results for filled buy trades are reported in Table~\ref{Tb:Estimation:FB}. First, the estimated parameters ($a$ and $\gamma$) of the PL model have obviously smaller standard errors than the estimated parameters ($c$ and $d$) of the LG model, no matter for the stock and the warrant. Comparing with the stock, the standard errors for the warrant are smaller due to the larger sample size. Second, the estimates of the PL model are more stable than those of the LG model. Taking the stock for example, the 11 estimates of $a$ lie in $[1.17, 1.57]$ and $\gamma$ lie in $[0.57, 0.84]$, while the 11 estimates of $c$ lie in $[4.56, 25.01]$ and $d$ lie in $[0.11, 1.55]$. The results are similar for the warrant. The 11 estimates of $a$ lie in $[1.30, 1.43]$ and $\gamma$ lie in $[0.52, 0.68]$, while the 11 estimates of $c$ lie in $[3.10, 6.62]$ and $d$ lie in $[0.62, 2.35]$. Third, the goodness of fit $R^2$ and the Mean Squared Error (MSE) show that both models have high fitting ability. However, we cannot conclude which model is better because the $R^2$ and the MSE of the 11 samples show inconsistent ranking. The estimation results for filled sell trades are totally similar and we report them in Table~\ref{Tb:Estimation:FS}.

\setlength\tabcolsep{1pt}
\begin{table*}[!htb]
  \centering
  \tbl{Results of parameter estimation for filled buy trades using two price impact models}
  {\begin{tabular*}{1\textwidth}{@{\extracolsep{\fill}}lccccccccccc}
    \hline\hline
    & 1 & 2 & 3 & 4 & 5 & 6 & 7 & 8 & 9 & 10 & 11 \\
    \hline
    \multicolumn{12}{l}{\textit{Power-estimated: Stock}}\\
     a &  1.170 &  1.571 &  1.378 &  1.395 &  1.207 &  1.510 &  1.337 &  1.263 &  1.355 &  1.299 &  1.394 \\
       & (0.128) & (0.375) & (0.186) & (0.086) & (0.130) & (0.173) & (0.106) & (0.103) & (0.054) & (0.076) & (0.086) \\
     $\gamma$ &  0.842 &  0.574 &  0.663 &  0.750 &  0.811 &  0.611 &  0.720 &  0.786 &  0.698 &  0.732 &  0.662 \\
       & (0.051) & (0.129) & (0.073) & (0.029) & (0.055) & (0.057) & (0.040) & (0.040) & (0.020) & (0.030) & (0.031) \\
     $R^2$ &  0.989 &  0.760 &  0.929 &  0.995 &  0.980 &  0.951 &  0.985 &  0.989 &  0.996 &  0.991 &  0.987 \\
     MSE &  0.049 &  0.766 &  0.182 &  0.024 &  0.070 &  0.139 &  0.047 &  0.040 &  0.012 &  0.026 &  0.034 \\
    \hline\multicolumn{12}{l}{\textit{Logarithm-estimated: Stock}}\\
     c & 25.007 &  4.559 &  6.465 & 11.535 & 15.532 &  5.559 &  9.329 & 14.262 &  8.780 &  9.762 &  7.013 \\
       & (12.532) & (1.824) & (1.527) & (1.007) & (5.054) & (0.918) & (1.513) & (3.353) & (1.108) & (1.442) & (0.801) \\
     d &  0.110 &  1.548 &  0.726 &  0.361 &  0.202 &  1.066 &  0.424 &  0.234 &  0.454 &  0.381 &  0.649 \\
       & (0.075) & (1.313) & (0.317) & (0.056) & (0.097) & (0.369) & (0.120) & (0.085) & (0.103) & (0.095) & (0.141) \\
     $R^2$ &  0.987 &  0.812 &  0.946 &  0.997 &  0.979 &  0.964 &  0.986 &  0.987 &  0.991 &  0.989 &  0.989 \\
     MSE &  0.061 &  0.599 &  0.138 &  0.013 &  0.074 &  0.100 &  0.045 &  0.048 &  0.028 &  0.031 &  0.031 \\
    \hline\multicolumn{12}{l}{\textit{Power-estimated: Warrant}}\\
     a &  1.404 &  1.325 &  1.391 &  1.420 &  1.432 &  1.414 &  1.339 &  1.302 &  1.297 &  1.323 &  1.340 \\
       & (0.044) & (0.029) & (0.048) & (0.050) & (0.058) & (0.054) & (0.035) & (0.023) & (0.022) & (0.045) & (0.027) \\
     $\gamma$ &  0.518 &  0.650 &  0.548 &  0.517 &  0.522 &  0.551 &  0.627 &  0.675 &  0.675 &  0.636 &  0.626 \\
       & (0.018) & (0.012) & (0.020) & (0.020) & (0.023) & (0.021) & (0.015) & (0.010) & (0.010) & (0.020) & (0.012) \\
     $R^2$ &  0.990 &  0.998 &  0.990 &  0.988 &  0.985 &  0.989 &  0.996 &  0.999 &  0.999 &  0.993 &  0.998 \\
     MSE &  0.012 &  0.004 &  0.015 &  0.016 &  0.022 &  0.018 &  0.007 &  0.003 &  0.003 &  0.012 &  0.004 \\
    \hline\multicolumn{12}{l}{\textit{Logarithm-estimated: Warrant}}\\
     c &  3.103 &  5.912 &  3.519 &  3.087 &  3.224 &  3.725 &  5.188 &  6.616 &  6.581 &  5.265 &  5.193 \\
       & (0.211) & (0.448) & (0.105) & (0.109) & (0.063) & (0.237) & (0.269) & (0.372) & (0.399) & (0.208) & (0.245) \\
     d &  2.226 &  0.729 &  1.772 &  2.351 &  2.238 &  1.690 &  0.913 &  0.616 &  0.618 &  0.871 &  0.900 \\
       & (0.327) & (0.099) & (0.110) & (0.181) & (0.096) & (0.227) & (0.089) & (0.060) & (0.065) & (0.063) & (0.079) \\
     $R^2$ &  0.990 &  0.995 &  0.998 &  0.997 &  0.999 &  0.992 &  0.997 &  0.997 &  0.997 &  0.998 &  0.997 \\
     MSE &  0.012 &  0.010 &  0.002 &  0.004 &  0.001 &  0.012 &  0.006 &  0.005 &  0.006 &  0.003 &  0.005 \\
    \hline\hline
  \end{tabular*}}
  \label{Tb:Estimation:FB}
  \raggedright \small Note: The table shows the estimated parameters for two models and two trading assets. Robust standard errors are reported in parentheses. MSE stands for Mean Squared Error.
\end{table*}

\begin{table*}[!htb]
  \centering
  \tbl{Results of parameter estimation for filled sell trades using two price impact models}
  {\begin{tabular*}{1\textwidth}{@{\extracolsep{\fill}}lccccccccccc}
    \hline\hline
    & 1 & 2 & 3 & 4 & 5 & 6 & 7 & 8 & 9 & 10 & 11 \\
        \cline{2-12}
    \multicolumn{12}{l}{\textit{Power-estimated: Stock}}\\
     a &  1.444 &  1.667 &  1.482 &  1.580 &  1.346 &  1.435 &  1.286 &  1.401 &  1.294 &  1.348 &  1.424 \\
       & (0.156) & (0.230) & (0.143) & (0.149) & (0.138) & (0.113) & (0.084) & (0.129) & (0.051) & (0.050) & (0.097) \\
     $\gamma$ &  0.680 &  0.586 &  0.680 &  0.584 &  0.756 &  0.653 &  0.787 &  0.708 &  0.728 &  0.678 &  0.696 \\
       & (0.050) & (0.066) & (0.047) & (0.046) & (0.053) & (0.040) & (0.032) & (0.046) & (0.020) & (0.019) & (0.034) \\
     $R^2$ &  0.978 &  0.934 &  0.977 &  0.962 &  0.977 &  0.978 &  0.994 &  0.980 &  0.996 &  0.995 &  0.988 \\
     MSE &  0.082 &  0.229 &  0.084 &  0.106 &  0.083 &  0.061 &  0.026 &  0.070 &  0.012 &  0.012 &  0.042 \\
    \hline\multicolumn{12}{l}{\textit{Logarithm-estimated: Stock}}\\
     c &  8.466 &  5.232 &  7.691 &  4.953 & 10.604 &  6.567 & 13.949 &  8.716 &  9.567 &  7.334 &  8.028 \\
       & (1.739) & (0.815) & (0.875) & (0.478) & (1.879) & (0.731) & (2.156) & (1.248) & (1.067) & (0.663) & (0.627) \\
     d &  0.538 &  1.473 &  0.666 &  1.414 &  0.378 &  0.758 &  0.251 &  0.510 &  0.387 &  0.577 &  0.583 \\
       & (0.217) & (0.519) & (0.147) & (0.297) & (0.112) & (0.163) & (0.061) & (0.134) & (0.073) & (0.095) & (0.084) \\
     $R^2$ &  0.975 &  0.964 &  0.989 &  0.985 &  0.985 &  0.988 &  0.994 &  0.987 &  0.994 &  0.993 &  0.995 \\
     MSE &  0.094 &  0.126 &  0.039 &  0.042 &  0.053 &  0.034 &  0.024 &  0.047 &  0.018 &  0.017 &  0.015 \\
     \hline\multicolumn{12}{l}{\textit{Power-estimated: Warrant}}\\
     a &  1.500 &  1.343 &  1.380 &  1.404 &  1.425 &  1.435 &  1.346 &  1.301 &  1.289 &  1.312 &  1.339 \\
       & (0.051) & (0.042) & (0.040) & (0.041) & (0.048) & (0.046) & (0.031) & (0.036) & (0.021) & (0.035) & (0.027) \\
     $\gamma$ &  0.518 &  0.635 &  0.559 &  0.537 &  0.526 &  0.533 &  0.615 &  0.668 &  0.664 &  0.636 &  0.616 \\
       & (0.019) & (0.018) & (0.017) & (0.017) & (0.019) & (0.018) & (0.014) & (0.016) & (0.010) & (0.016) & (0.012) \\
     $R^2$ &  0.991 &  0.995 &  0.993 &  0.992 &  0.990 &  0.991 &  0.997 &  0.996 &  0.999 &  0.996 &  0.998 \\
     MSE &  0.015 &  0.010 &  0.010 &  0.010 &  0.014 &  0.013 &  0.006 &  0.007 &  0.002 &  0.007 &  0.004 \\
    \hline\multicolumn{12}{l}{\textit{Logarithm-estimated: Warrant}}\\
     c &  3.362 &  5.503 &  3.683 &  3.389 &  3.240 &  3.419 &  4.876 &  6.225 &  6.055 &  5.237 &  4.875 \\
       & (0.128) & (0.456) & (0.124) & (0.132) & (0.125) & (0.124) & (0.223) & (0.438) & (0.405) & (0.290) & (0.268) \\
     d &  2.223 &  0.823 &  1.608 &  1.921 &  2.168 &  1.981 &  0.998 &  0.665 &  0.679 &  0.857 &  0.979 \\
       & (0.190) & (0.125) & (0.110) & (0.159) & (0.181) & (0.155) & (0.086) & (0.082) & (0.079) & (0.086) & (0.101) \\
     $R^2$ &  0.997 &  0.993 &  0.998 &  0.997 &  0.997 &  0.997 &  0.997 &  0.996 &  0.996 &  0.997 &  0.996 \\
     MSE &  0.005 &  0.013 &  0.003 &  0.004 &  0.004 &  0.004 &  0.004 &  0.008 &  0.007 &  0.006 &  0.006 \\
    \hline\hline
  \end{tabular*}}
  \label{Tb:Estimation:FS}
  \raggedright \small Note: The table shows the estimated parameters for two models and two trading assets. Robust standard errors are reported in parentheses. MSE stands for Mean Squared Error.
\end{table*}

\section{Out-of-sample predictive accuracy}

\begin{figure*}[!htb]
  \centering
  \includegraphics[width=0.38\linewidth]{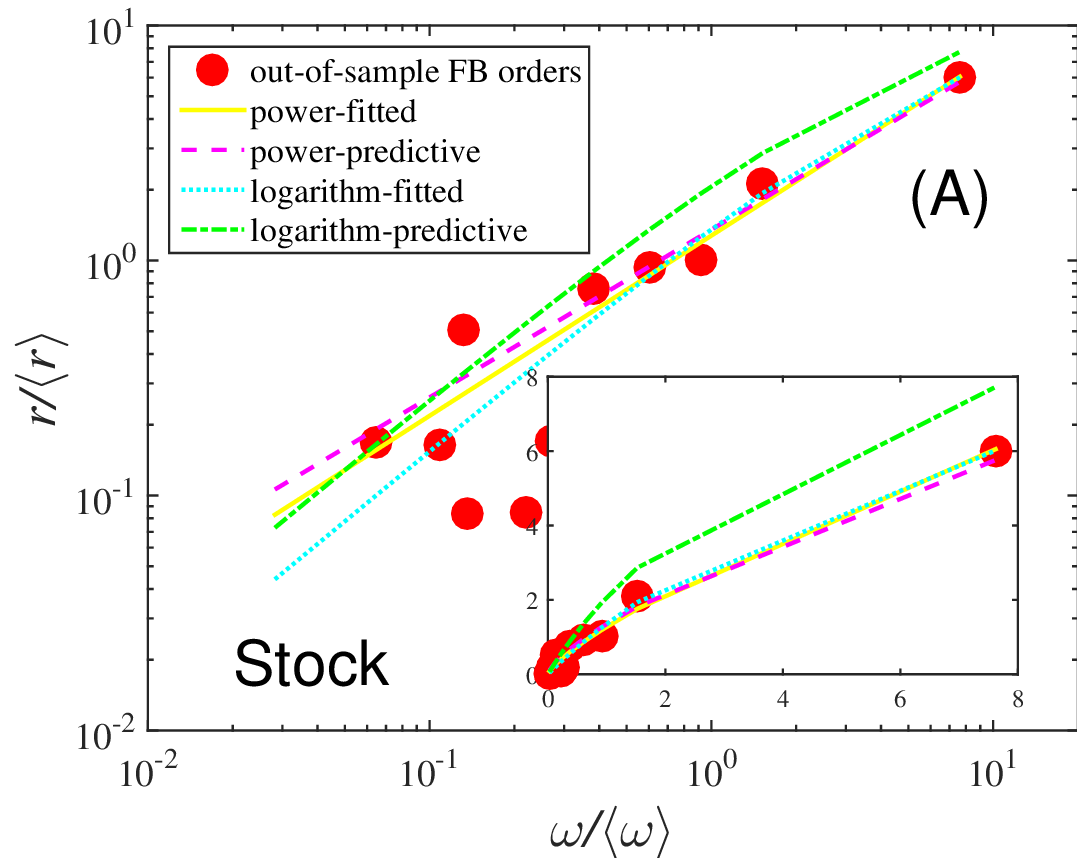}
  \includegraphics[width=0.38\linewidth]{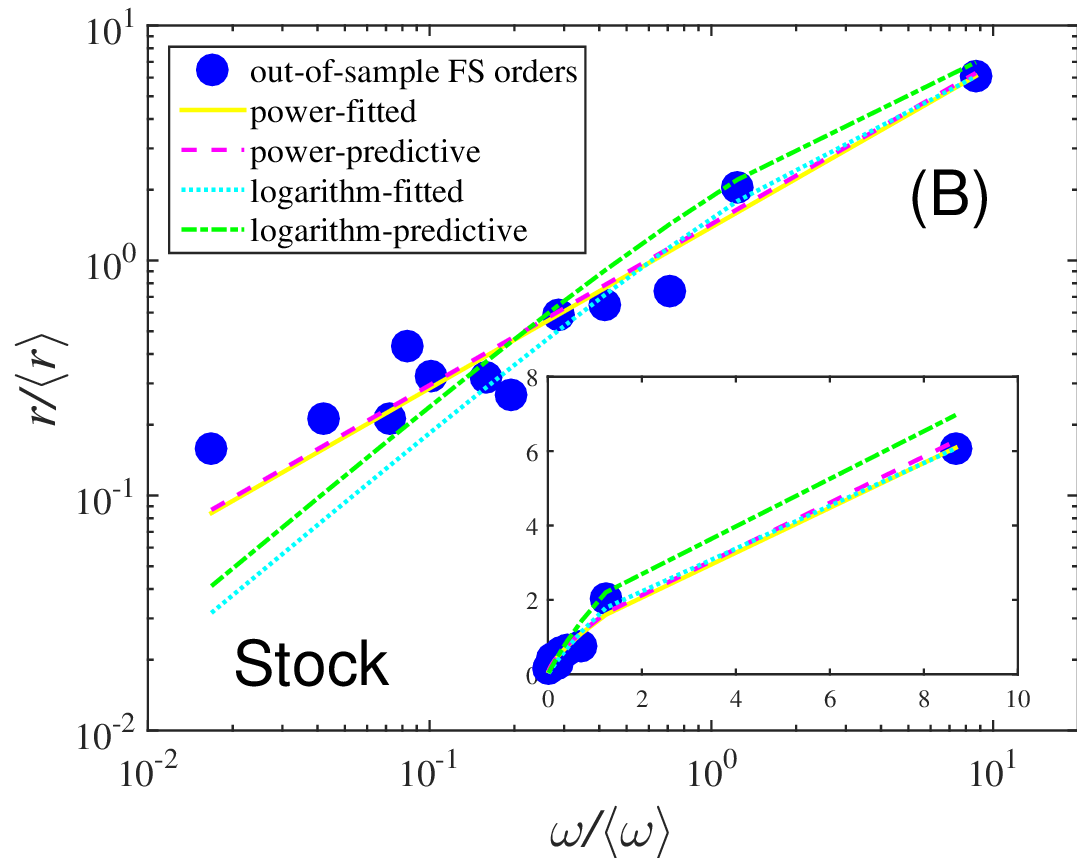}\\
  \vskip  +0.02\textwidth
  \includegraphics[width=0.38\linewidth]{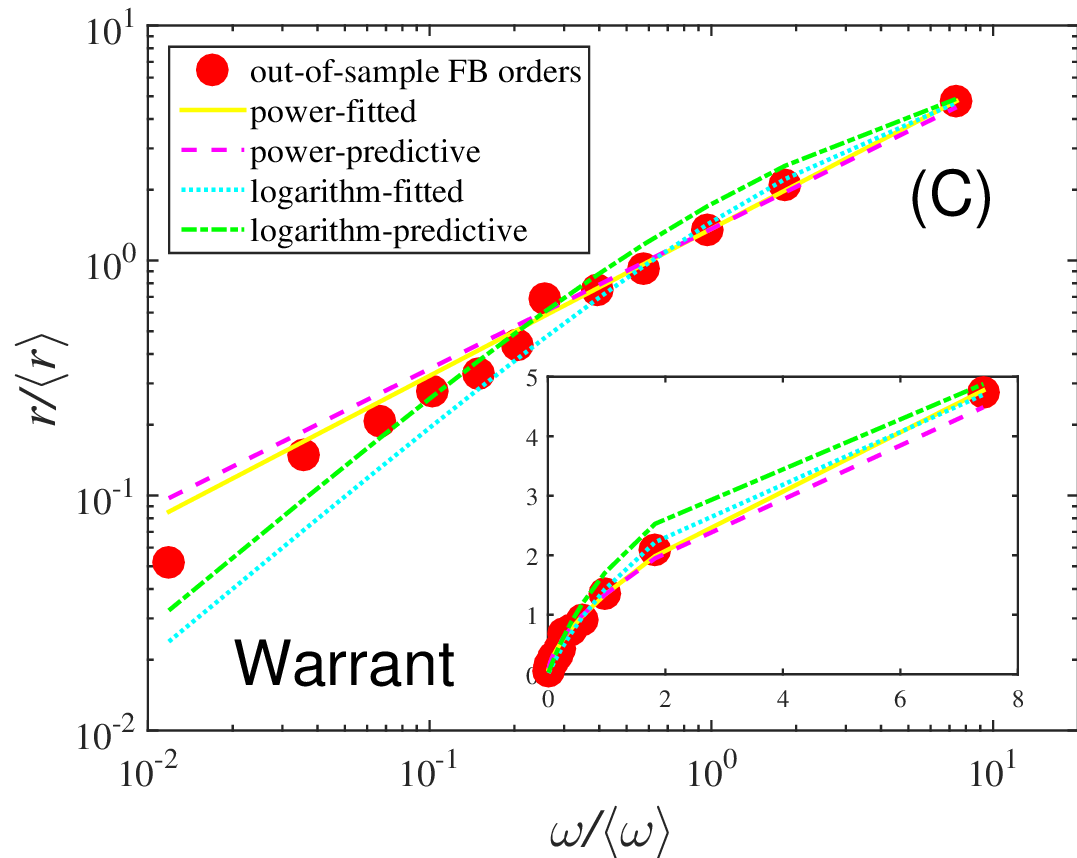}
  \includegraphics[width=0.38\linewidth]{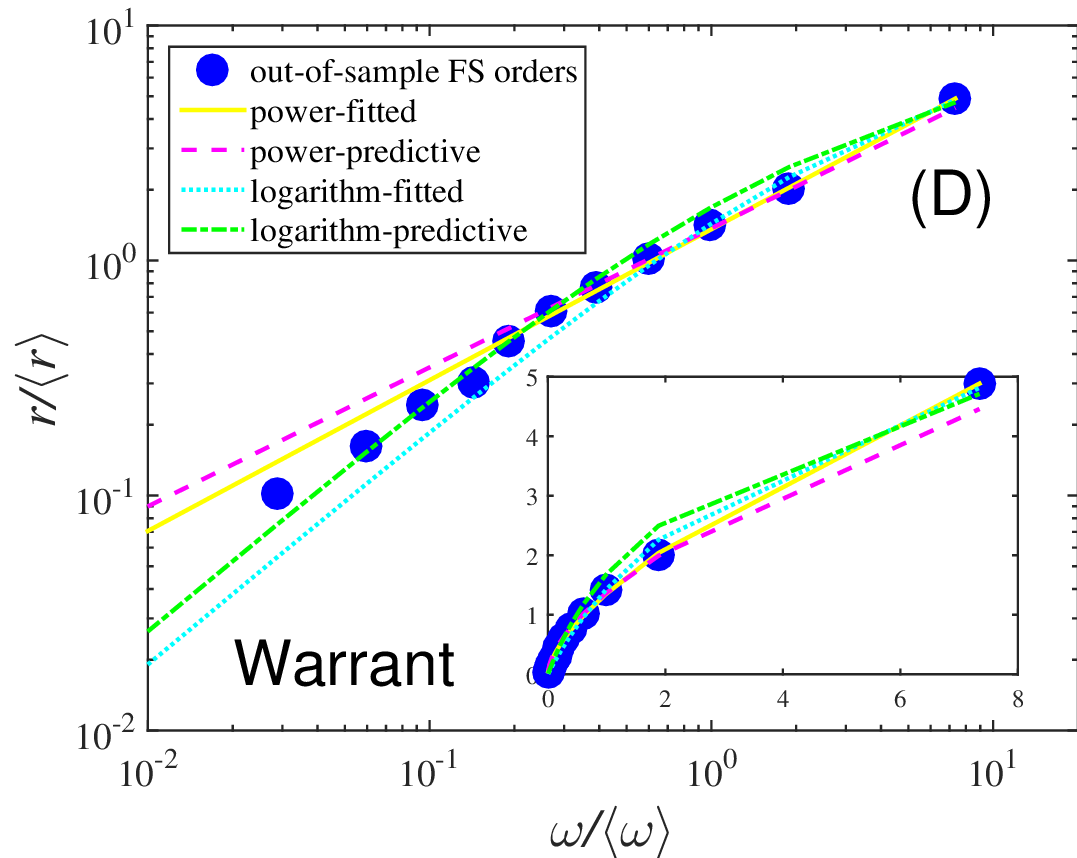}
  \caption{Out-of-sample prediction for filled trades of Stock 600019 and Warrant 580000. The magenta line draws the power-law prediction based on the parameters $a$ and $\gamma$, which are mean values of the first 11 months' estimated values. The yellow line represents the real power-law fitted curve using out-of-sample trades. The green line draws the logarithmic prediction based on the parameters $c$ and $d$, which are mean values of the first 11 months' estimated values. The cyan line represents the real logarithmic fitted curve using out-of-sample trades. The embedded graph is the same as the main graph except for its linear scale coordinates.  (A) Stock: FB trades. (B) Stock: FS trades. (C) Warrant: FB trades. (D) Warrant: FS trades. }\label{Fig:IPI:Outsample}
\end{figure*}

Considering that the estimated parameters in the first 11 months have no trends but fluctuate in relatively stable intervals, we use their mean values to test the out-of-sample performance of the two models. Fig.~\ref{Fig:IPI:Outsample} illustrates the case of splitting groups $M=12$ for both FB and FS trades of the stock and FB and FS trades of the warrant. To check for robustness, Table~\ref{Tb:outsample} reports the results for three cases of $M=12,15,18$. In Fig.~\ref{Fig:IPI:Outsample}, we not only draw the predictive price impact curves, but also draw the fitted curves using trades of the 12th month. It is reasonable that the predictive results are no better than the fitted ones and thus deviate from the real pairs ($\omega/\langle \omega \rangle$, $r/\langle r \rangle$) more. This can be seen in Fig.~\ref{Fig:IPI:Outsample}. More importantly, in log-log scale coordinates, no matter for the PL model or the LG model, all the predictive curves show some small deviations from the real price impacts when the trade sizes $\omega/\langle \omega \rangle$ are small. However, the predictive curves of the LG model show obvious deviations when the trade sizes $\omega/\langle \omega \rangle$ are large, while the PL model curves do not. This can be observed more clearly in the linear scale view (the inset). Therefore, we infer that the PL model provides more precise out-of-sample forecasting.

\begin{table*}[!htb]
  \centering
  \tbl{Out-of-sample predictive accuracy comparisons, using Mean Squared Error criterion.}
  {\begin{tabular*}{1\textwidth}{@{\extracolsep{\fill}}lcccc|cccc|cccc}
    \hline\hline
     &  \multicolumn{2}{c}{Stock$^{(1)}$} & \multicolumn{2}{c}{Warrant$^{(1)}$} & \multicolumn{2}{c}{Stock$^{(2)}$} & \multicolumn{2}{c}{Warrant$^{(2)}$} &  \multicolumn{2}{c}{Stock$^{(3)}$} & \multicolumn{2}{c}{Warrant$^{(3)}$} \\
        \cline{2-13}
     & FB & FS & FB & FS & FB & FS & FB & FS & FB & FS & FB & FS\\
     \hline
    \multicolumn{4}{l}{\textit{Power-predictive}}&&&&&&&&\\
     a &  1.353 &  1.428 &  1.362 &  1.370 & 1.365 &  1.437 &  1.371 &  1.378 &  1.386 &  1.443 &  1.379 &  1.384 \\
     $\gamma$ &  0.714 &  0.685 &  0.595 &  0.592 &  0.717 &  0.696 &  0.598 &  0.594 &  0.714 &  0.699 &  0.597 &  0.595 \\
     MSE &  0.061 &  0.045 &  0.013 &  0.022 &  0.095 &  0.157 &  0.013 &  0.028 &  0.127 &  0.087 &  0.015 &  0.039 \\
     \hline
    \multicolumn{4}{l}{\textit{Logarithm-predictive}}&&&&&&&&&\\
     c & 10.709 &  8.282 &  4.674 &  4.533 & 13.563 &  9.951 &  5.131 &  4.992 & 12.423 & 11.279 &  5.479 &  5.295 \\
     d &  0.560 &  0.685 &  1.357 &  1.355 &  0.456 &  0.529 &  1.198 &  1.203 &  0.394 &  0.446 &  1.116 &  1.114 \\
     MSE &  0.511 &  0.152 &  0.043 &  0.035 &  0.833 &  0.259 &  0.052 &  0.053 &  0.214 &  0.233 &  0.073 &  0.071 \\
    \hline\hline
  \end{tabular*}}
  \label{Tb:outsample}
  \raggedright \small Note: (1), (2) and (3) stand for the cases of $M=12$, $M=15$ and $M=18$ respectively.
\end{table*}

Table~\ref{Tb:outsample} confirms the above inference. We can observe that, no matter for the stock or the warrant, and no matter for the FB trades or the FS trades, the out-of-sample MSEs of the PL model are significantly smaller than those of the LG model. Alternative $M$ values do not change the results. Therefore we can conclude that the PL model performs better in the our-of-sample tests.

\section{Price impact correlations}

\begin{figure}[!htb]
  \centering
  \includegraphics[width=0.45\linewidth]{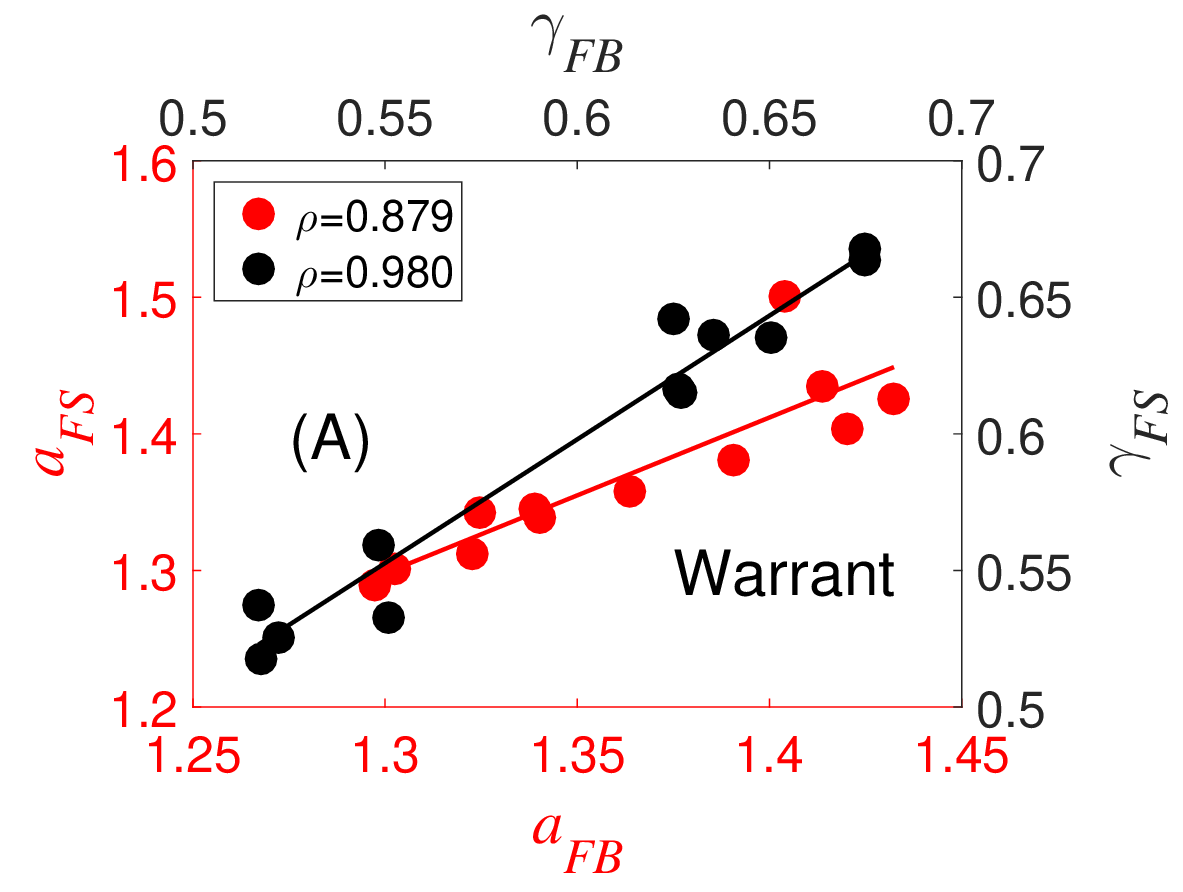}
  \hskip  +0.005\textwidth
  \includegraphics[width=0.45\linewidth]{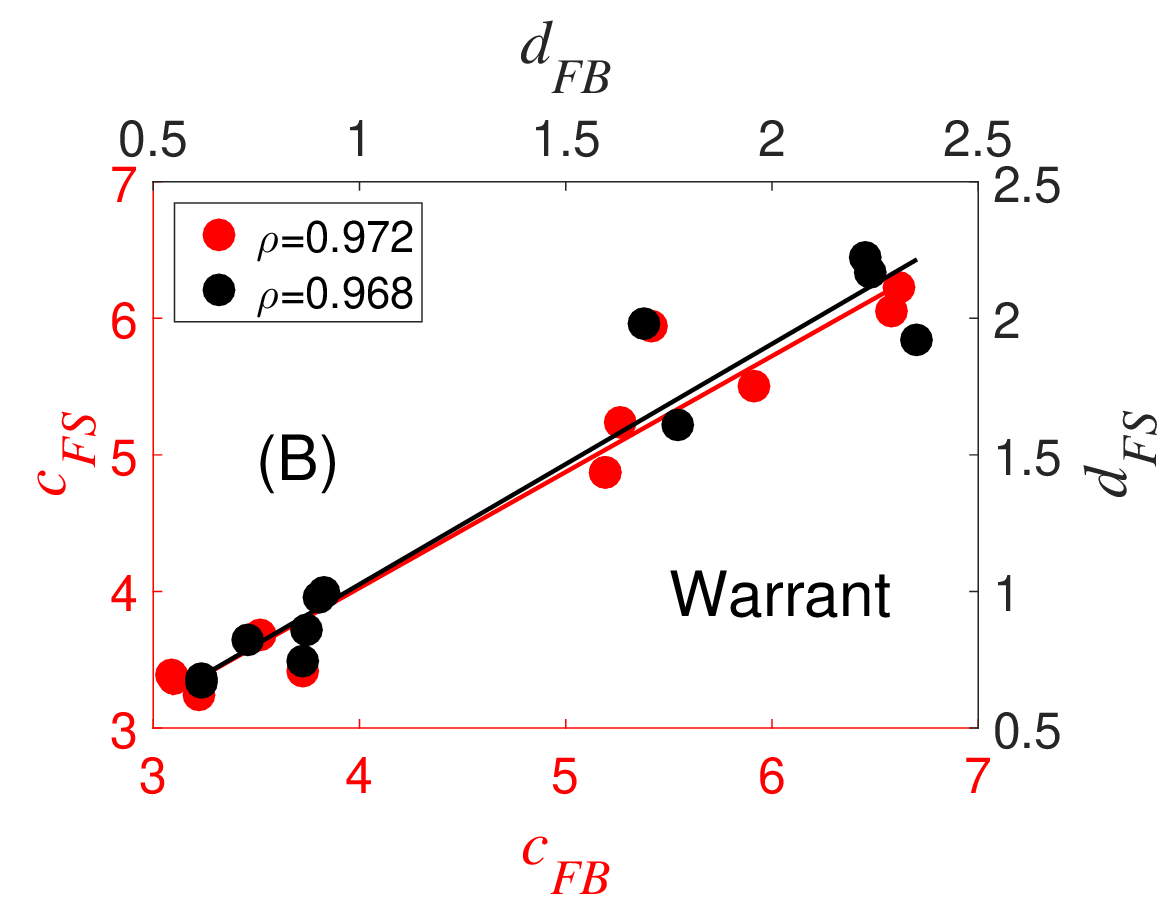}\\
  \includegraphics[width=0.45\linewidth]{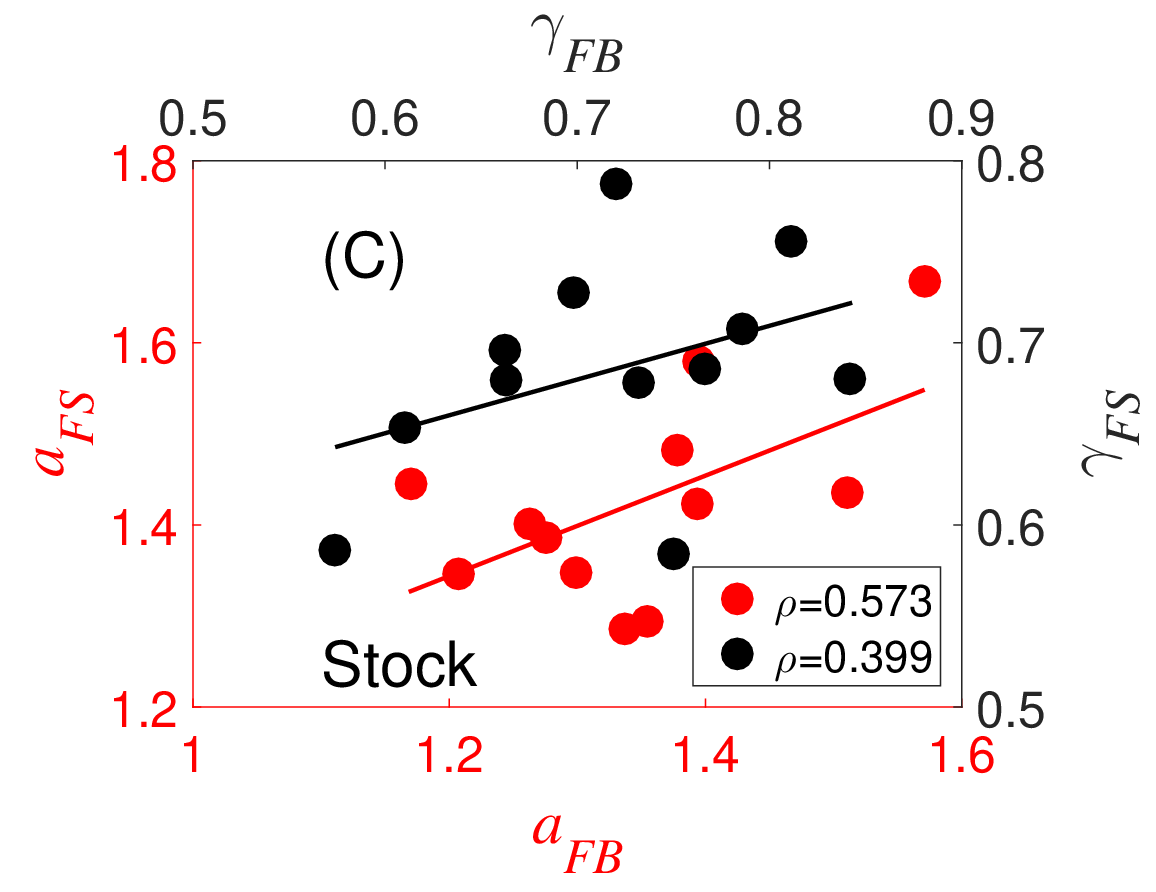}
  \hskip  +0.005\textwidth
  \includegraphics[width=0.45\linewidth]{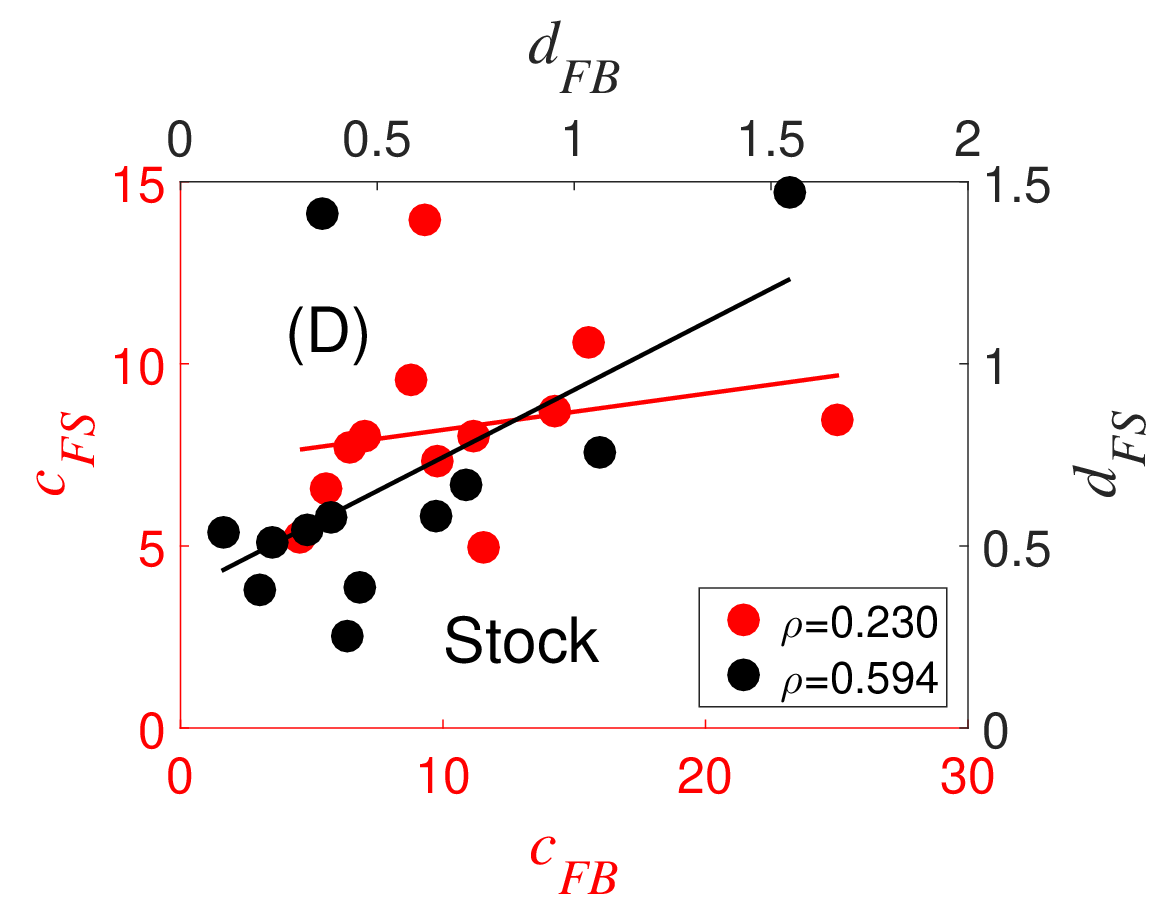}
  \caption{Parameter correlation between FB trades and FS trades using two price impact model. (A)For Warrant, power-law model. (B)For Warrant, logarithmic model. (C)For Stock, power-law model. (D)For Stock, logarithmic model.}\label{Fig:IPI:BS:Correlation}
\end{figure}

\begin{figure}
  \centering
  \includegraphics[width=0.45\linewidth]{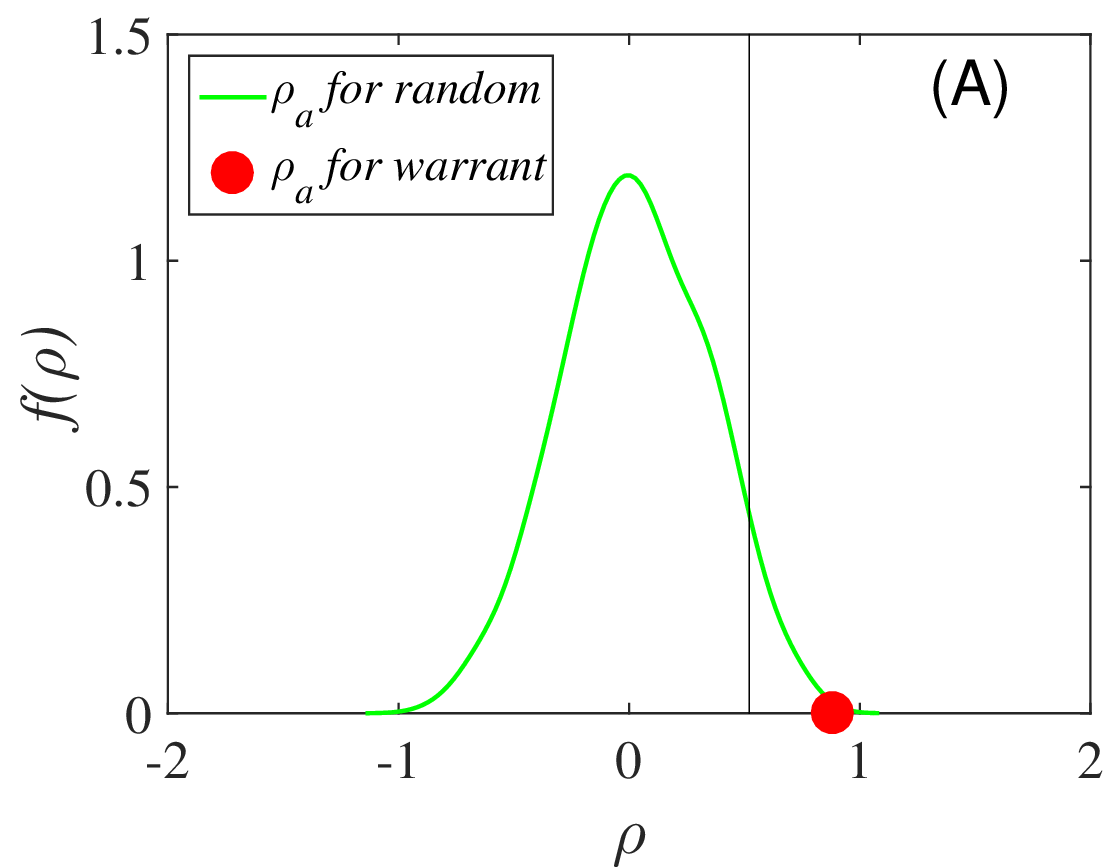}
  \includegraphics[width=0.45\linewidth]{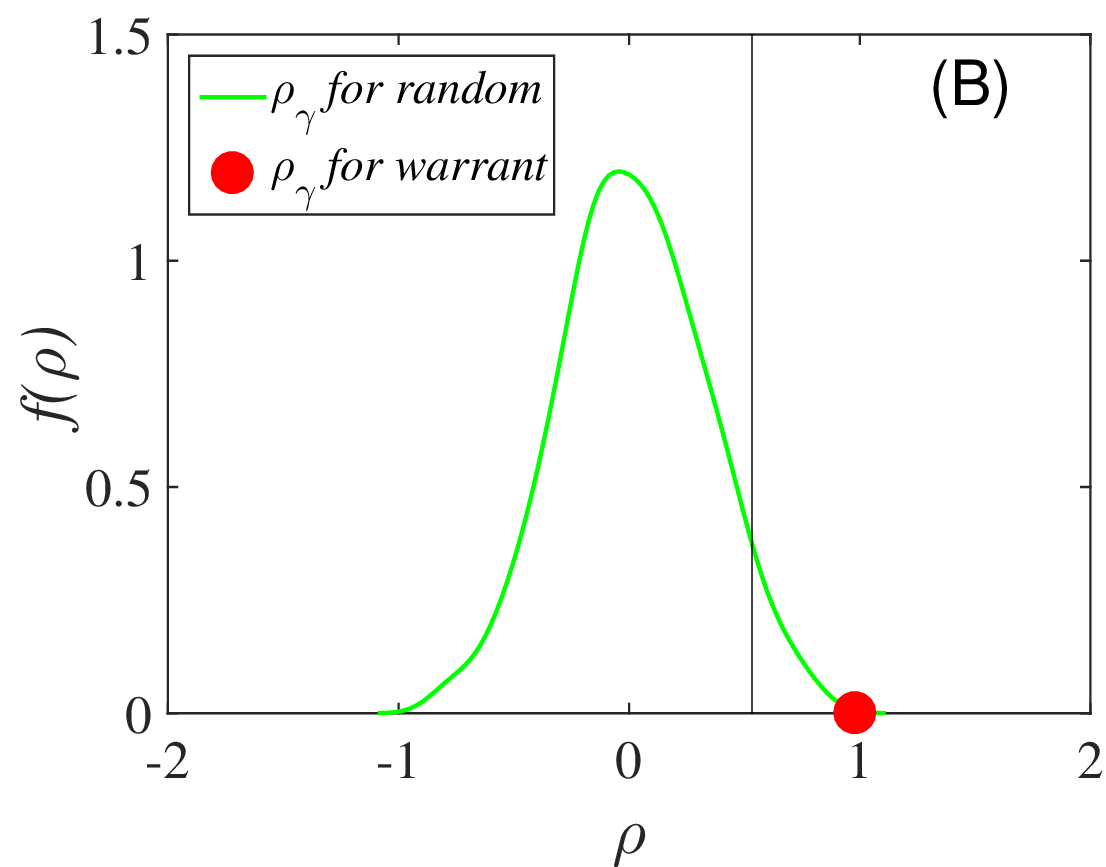}\\
  \vskip  +0.01\textwidth
  \includegraphics[width=0.45\linewidth]{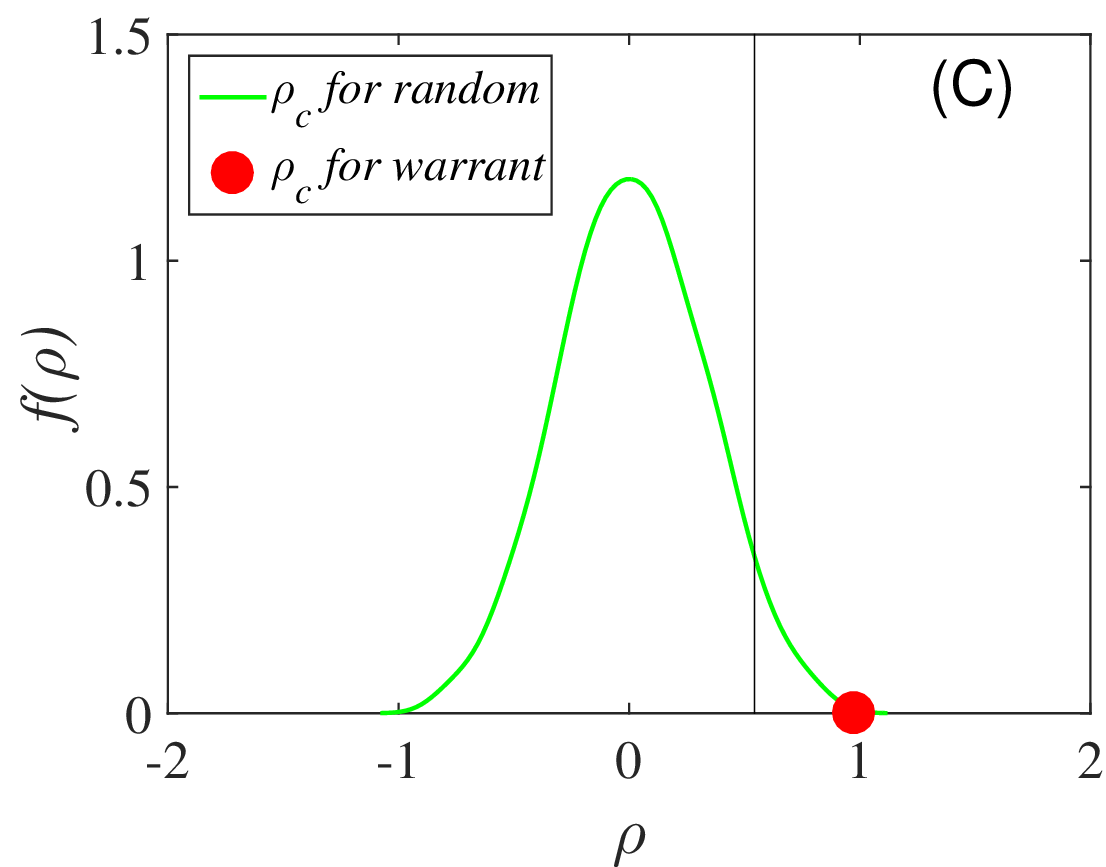}
  \includegraphics[width=0.45\linewidth]{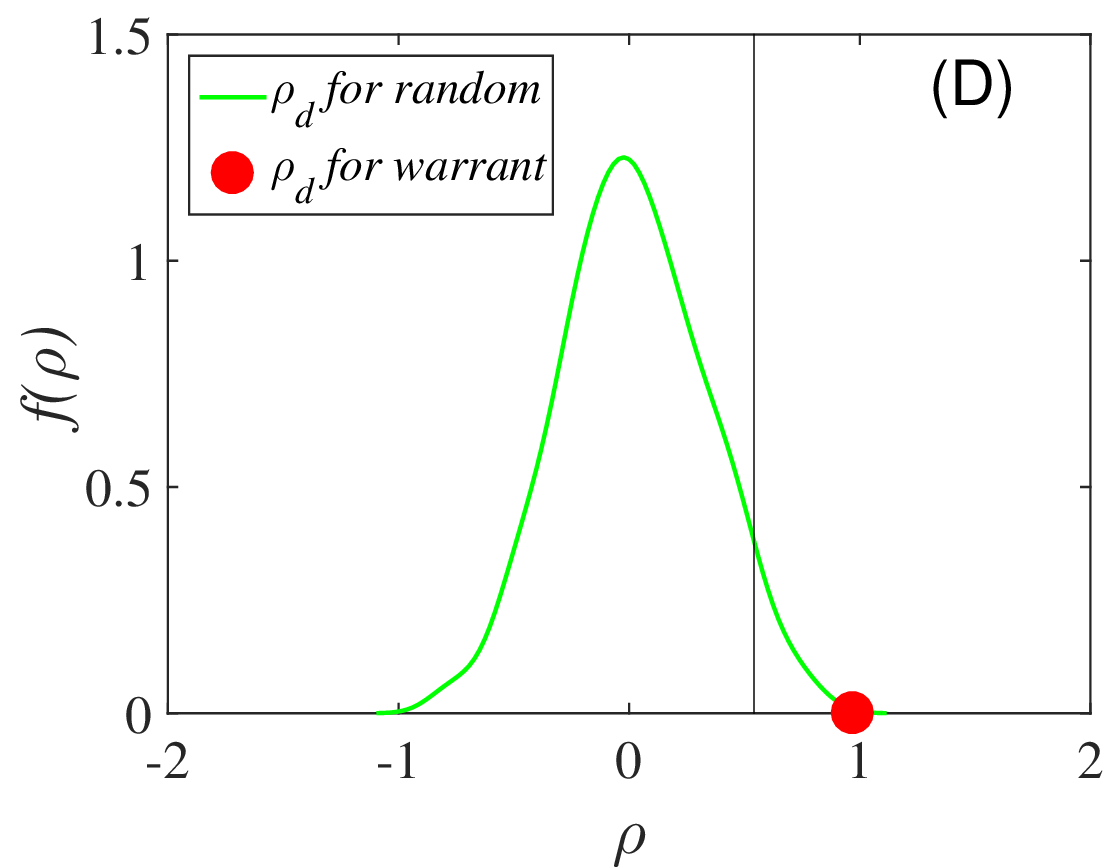}\\
  \vskip  +0.01\textwidth
  \includegraphics[width=0.45\linewidth]{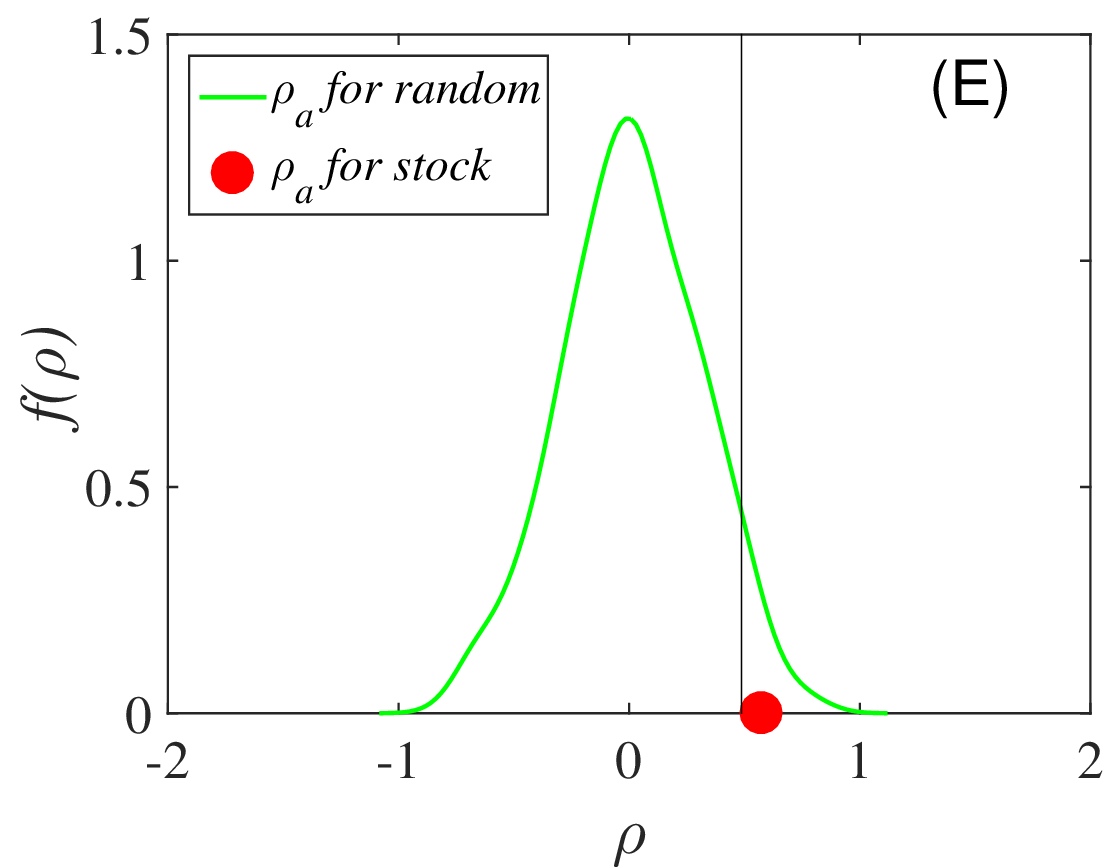}
  \includegraphics[width=0.45\linewidth]{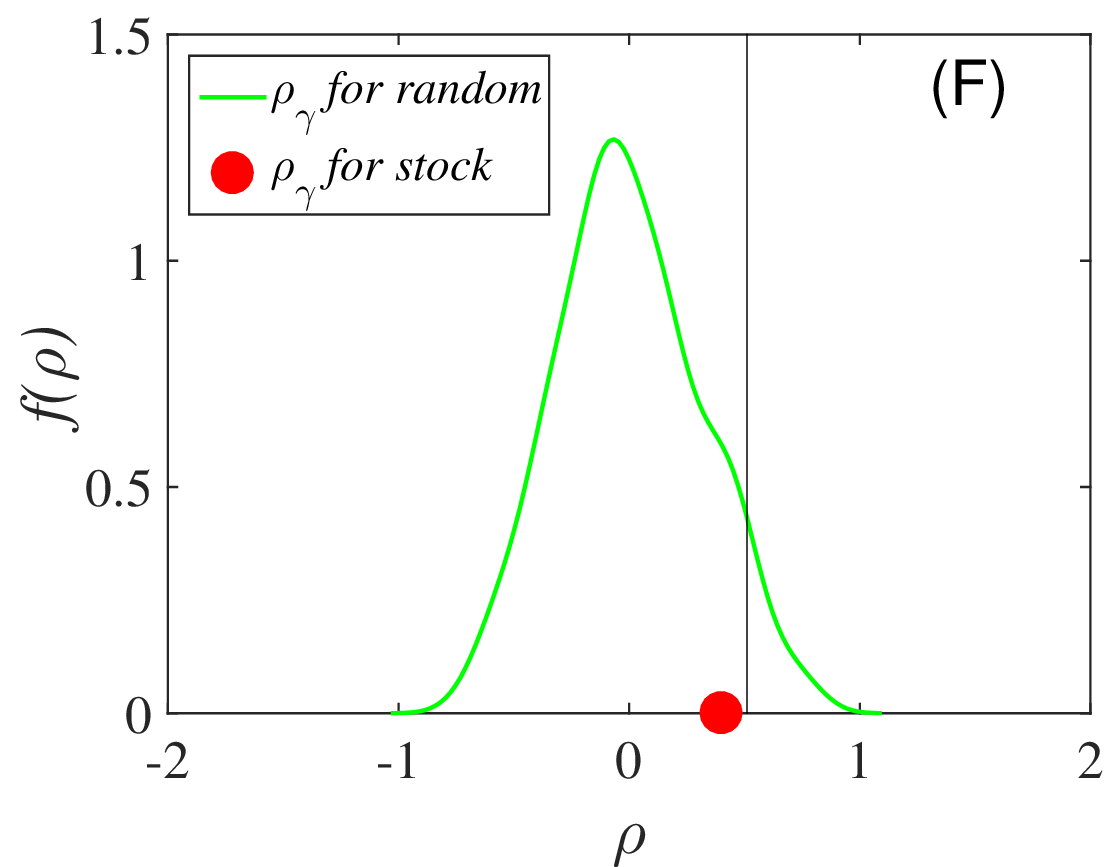}\\
  \vskip  +0.01\textwidth
  \includegraphics[width=0.45\linewidth]{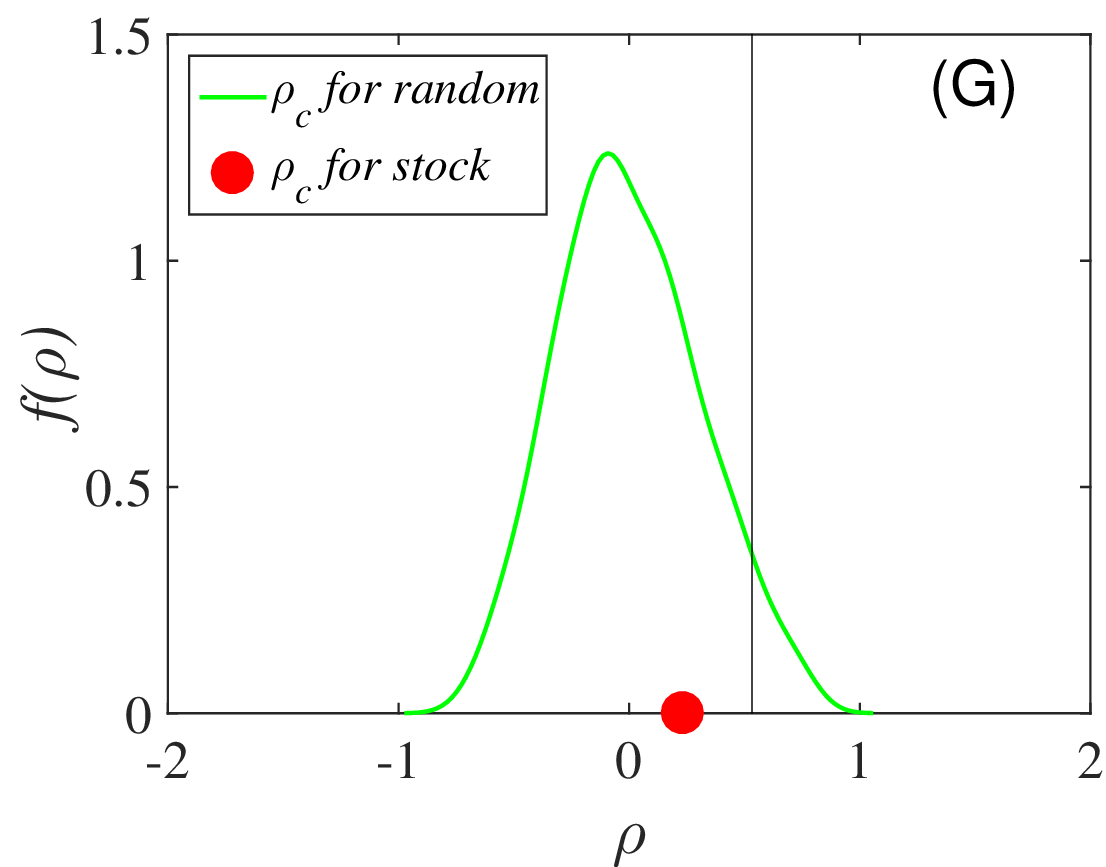}
  \includegraphics[width=0.45\linewidth]{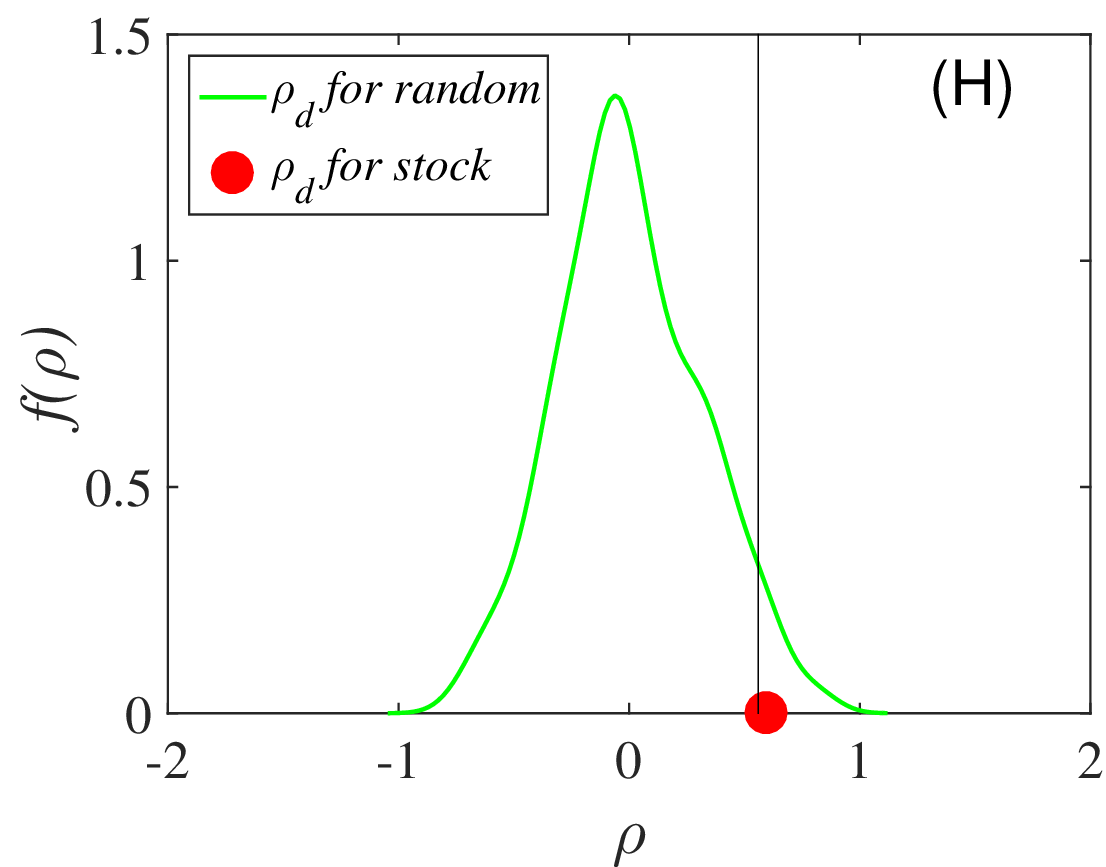}
  \caption{Statistical significance of parameter correlation between FB trades and FS trades using two price impact model. In each plot, the red point represents the parameter correlation obtained by original sequential order flow; the green line represents the probability distribution of parameter correlation for 500 random reshuffled order flow series; the vertical black line represents the quantile of 5\%. (A-B)For Warrant, power-law model. (C-D)For Warrant, logarithmic model. (E-F)For Stock, power-law model. (G-H)For Stock, logarithmic model.}\label{Fig:IPI:BS:Correlation:Shuffle}
\end{figure}

In Fig.~\ref{Fig:IPI:Insample}, we observe that in general, there is no asymmetry between FB trades and FS trades. In this part, we will examine whether the price impacts of the FB trades change simultaneously with those of the FS trades. Fig.~\ref{Fig:IPI:BS:Correlation} shows that, no matter which price impact model we use, the estimated parameters ($a$, $\gamma$, $c$, $d$) of the FB trades are positively correlated with those of the FS trades. Especially for the warrant, there are significant positive correlations ($\rho>0.87$) between them.

In order to confirm that these positive serial correlations are not accidental, we shuffle FB trade series and FS trade series for 500 times and do the same analysis. This allow us to construct the distribution of parameter correlations of 500 FB-FS random series pairs. In Fig.~\ref{Fig:IPI:BS:Correlation:Shuffle}, we compare the parameter correlations between original trade series and shuffled trade series. We can confirm that the positive serial correlations are very significant at 5\% level for warrant, thus the consistent behaviors of price impact between FB and FS trades are significant. For stock, the positive correlations can also confirmed even though they are not such obvious as warrant.

\section{Summary and discussions}

In summary, we utilize order matching data to examine two price impact models at the transaction level. We consider a stock and a warrant with different tick sizes and trading activities. The stock and the warrant behave similarly in comparison of the power-law model and the logarithmic model. In some months the power-law model provides better goodness of fit with lower Mean Squared Error, while in other months the logarithmic model does better. However, the estimated parameters of the power-law model behave more stable over time and have obviously smaller standard errors. Due to this advantage, the power-law model performs more accurate in out-of-sample forecasting. In addition, we find that positive correlations exist in the estimated parameters of price impact models between filled buy trades and filled sell trades. Our discoveries contribute to the transaction cost calculation and the optimal trade execution.

We notice that a market invariance model has been proposed recently,\cite{Kyle-Obizhaeva-2016-Em} which is a more complicated liquidity scaling. We will give a test to such alternatives in a further study.

\section*{Acknowledgements}

We acknowledge financial support from National Natural Science Foundation of China (71501072, 71532009) and the Fundamental Research Funds for the Central Universities.

\section*{References}

\bibliographystyle{ws-acs}
\bibliography{E:/Papers/Auxiliary/Bibliography}

%
%
%
%
%

\end{document}